\documentclass[letterpaper]{article} 
\usepackage{aaai24}  
\usepackage{times}  
\usepackage{helvet}  
\usepackage{courier}  
\usepackage[hyphens]{url}  
\usepackage{graphicx} 
\usepackage{natbib}  
\usepackage{caption} 
\usepackage{algorithm}
\usepackage{algorithmic}
\usepackage{booktabs} 

\usepackage{newfloat}
\usepackage{listings}
\DeclareCaptionStyle{ruled}{labelfont=normalfont,labelsep=colon,strut=off} 
\floatstyle{ruled}
\newfloat{listing}{tb}{lst}{}
\floatname{listing}{Listing}

\title{AI Data Transparency: an Exploration Through the Lens of AI Incidents}
\author{
    Sophia Worth$^1$, Ben Snaith$^1$, Arunav Das$^2$, Gefion Thuermer$^1$, Elena Simperl$^1$
}

\affiliations{
    $^1$Open Data Institute, London, UK\\

    $^2$King's College London, London, UK
}

\begin{document}

\maketitle

\begin{abstract}
Knowing more about the data used to build AI systems is critical for allowing different  stakeholders to play their part in ensuring responsible and appropriate deployment and use. Meanwhile, a 2023 report shows that data transparency lags significantly behind other areas of AI transparency in popular foundation models. In this research, we sought to build on these findings, exploring the status of public documentation about data practices within AI systems generating public concern. 

Our findings demonstrate that low data transparency persists across a wide range of systems, and further that issues of transparency and explainability at model- and system- level create barriers for investigating data transparency information to address public concerns about AI systems. We highlight a need to develop systematic ways of monitoring AI data transparency that account for the diversity of AI system types, and for such efforts to build on further understanding of the needs of those both supplying and using data transparency information. 
\end{abstract}

\section{Introduction}

Data is at the heart of AI systems, playing a key role across the AI life cycle \cite{sambasivan_everyone_2021} and in determining the “consequences in downstream AI deployment” \cite{jarrahi_principles_2023,sambasivan_everyone_2021}. For example, many are now aware of the serious risks associated with facial recognition systems developed on biased datasets \cite{buolamwini_gender_2018}, or AI systems revealing confidential information \cite{curzon_privacy_2021}. Such examples demonstrate that it is not only \textit{which} data is used to build AI systems that is important, but also how well those developing, deploying and using AI systems understand biases, limitations and legal obligations associated with use of this data, to ensure systems are implemented appropriately.

Transparency about the use of data in AI systems is rightly gaining attention \cite{hardinges_we_2023} as a pressing area for improvement in AI governance, in what researchers have called a “growing data transparency crisis” \cite{longpre_data_2023-1} given widespread difficulties in understanding the use of data across AI systems, and the ‘lineage’ of that data (who created it, how it has been curated, its limitations, etc.). 

The past several years have seen a significant range of AI data transparency efforts emerging across sectors. These have both focused on documentation practices to describe specific datasets that are commonly used in AI systems (including for example, Datasheets and Data Cards) and as part of model- and system- transparency approaches including Model or System Cards, which include information about how data has been used to build AI systems. However, evidence suggests that use of these documentation practices often remains ad-hoc and inconsistent \cite{yang_navigating_2024, liang_advances_2022}. One Microsoft study demonstrated that UX designers experience a lack of data transparency information to understand pre-trained models for responsibly implementing these in AI products \cite{liao_designerly_2023}, while the Stanford Foundation Model Transparency Index demonstrated that the data transparency of a selection of 10 popular foundation models is very low \cite{bommasani_foundation_2023}. More broadly, there remains limited further evidence on how far the range of data transparency approaches have permeated to help different stakeholders understand AI systems and the role of data in their downstream impacts.

At the same time, there are widespread and increasing demands from diverse stakeholders including the public, civil society, private sector organisations and governments for better AI transparency \cite{jobin_global_2019}, and AI data transparency more specifically \cite{hardinges_we_2023}.

In this research we sought to evaluate the landscape of AI data transparency among those AI systems generating public concern. To do so, we made use of the AI incidents database (AIID) generated by the Partnership on AI as a basis for identifying AI systems associated with public concern. Using an initial sample of 135 incidents that took place between January 2022 and March 2024, which generated a sample of 54 AI systems, we adopted elements of the search protocol developed by the creators of the Stanford Foundation Model Transparency Index (FMTI) \cite{bommasani_foundation_2023} to explore the data transparency of AI systems underlying the sample of incidents. We found that the low levels of AI data transparency identified in the Stanford research were replicated across a 25 models (a wider range than the 10 models explored in the FMTI), and further identified issues of transparency and explainability at model- and system- levels that create barriers for investigating data transparency information. We propose that further efforts to investigate and monitor the landscape of AI data transparency, including better understanding the needs of those using and supplying transparency information, will be important for helping those designing and deploying AI systems, as well as regulators and researchers on responsible AI to target their efforts, generating benefits in supporting wider improvements in AI data transparency.

\section{Background}

AI data transparency should be considered in relation to the ‘needs’ for transparency information. Historically, the data used to build AI systems has commonly been viewed as a ‘technical’ element, and therefore the details of data used only relevant to those with technical expertise \cite{jarrahi_principles_2023} - an increasingly disputed paradigm given the importance of data and its governance for the ‘downstream’ impacts of AI system use \cite{sambasivan_everyone_2021}. As a result, there is increasing demand for publicly documented information about how data is used within AI systems that is accessible to a wide audience, and that helps to understand system impacts - and AI data transparency is defined as such in the context of this research, following definitions used by \citet{bommasani_foundation_2023}and \citet{bertino_quest_2020}. 

In this section, we first explore the forms that AI data transparency can take, before highlighting the diverse needs for transparency information and the tensions this can generate, and finally the current evidence on the landscape of data transparency across diverse AI systems. 

\subsection{Multiple forms of AI data transparency}
When it comes to AI data transparency, much emphasis has been placed on approaches focused on specific datasets \cite{micheli_landscape_2023} such as ‘datasheets’, ‘data cards’ and ‘data nutrition labels’, which typically target developers/those building AI systems to make appropriate use of those datasets \cite{hallinan_data_2020, gebru_datasheets_2021, pushkarna_data_2022}. This focus is essential considering these stakeholders’ significant responsibility in addressing elements such as quality, bias, and contextual relevance of data. While they may complement one another, it is important to differentiate such transparency approaches from those that help a more general audience to understand which datasets and how these datasets have been implemented in AI systems (the focus of this research). For these forms of data transparency, we need to look at model- and system- level transparency approaches such as Model Cards \cite{mitchell_model_2019, oecd_model_2022}, System Cards \cite{meta_system_2022}, or perhaps the UK’s Algorithmic Transparency Reporting Standard \cite{cddo_algorithmic_2024}, and which data transparency information they incorporate. Importantly, it is often advised that developers incorporate data-level transparency approaches such as data cards into their model- and system- level transparency approaches, although we have not identified research exploring how often this is completed. 

\subsection{Who needs AI data transparency information?}

In a December 2023 public scandal, the generation of child sexual abuse material (CSAM) by the Stable Diffusion\footnote{Created by StabilityAI, a major generative AI company who have partnered with organisations such as Intel.} text-to-image model stemmed from the system being trained on the popular, vast and openly available LAION-5B training dataset, which contained over a thousand instances of CSAM material \cite{verma_exploitive_2023}. Stanford researchers investigating the scandal highlighted the issues raised by poor transparency about the data used to build AI systems, including for identifying in which other models the LAION-5B dataset has been used, how it has been used, and whether similar risks for exposure of this illegal material are likely to emerge \cite{thiel_identifying_2023}. Without such information, it is difficult for stakeholders all across the AI lifecycle to understand how to interpret AI systems, their component parts and their limitations, and how to use the technologies appropriately. 

While public concern about AI system deployment and the social and ethical implications are high \cite{modhvadia_how_2023}, particularly with regards to development and deployment among private companies, research has shown success in transparency information improving trust among users and deployers, with one study demonstrating in particular the effectiveness of ‘data-centric explanations’ for helping users to understand and evaluate machine learning systems \cite{anik_data-centric_2021}. Further, in the context of the movement towards responsible AI, researchers argue \cite{bommasani_foundation_2023-1} that beyond these practical needs, requiring system developers to work in the open increases scrutiny and accountability, thereby encouraging responsible practices, for example creating a need to demonstrate how harms are mitigated.

\begin{figure*} [h!]
    \centering
    \includegraphics[scale =0.24]{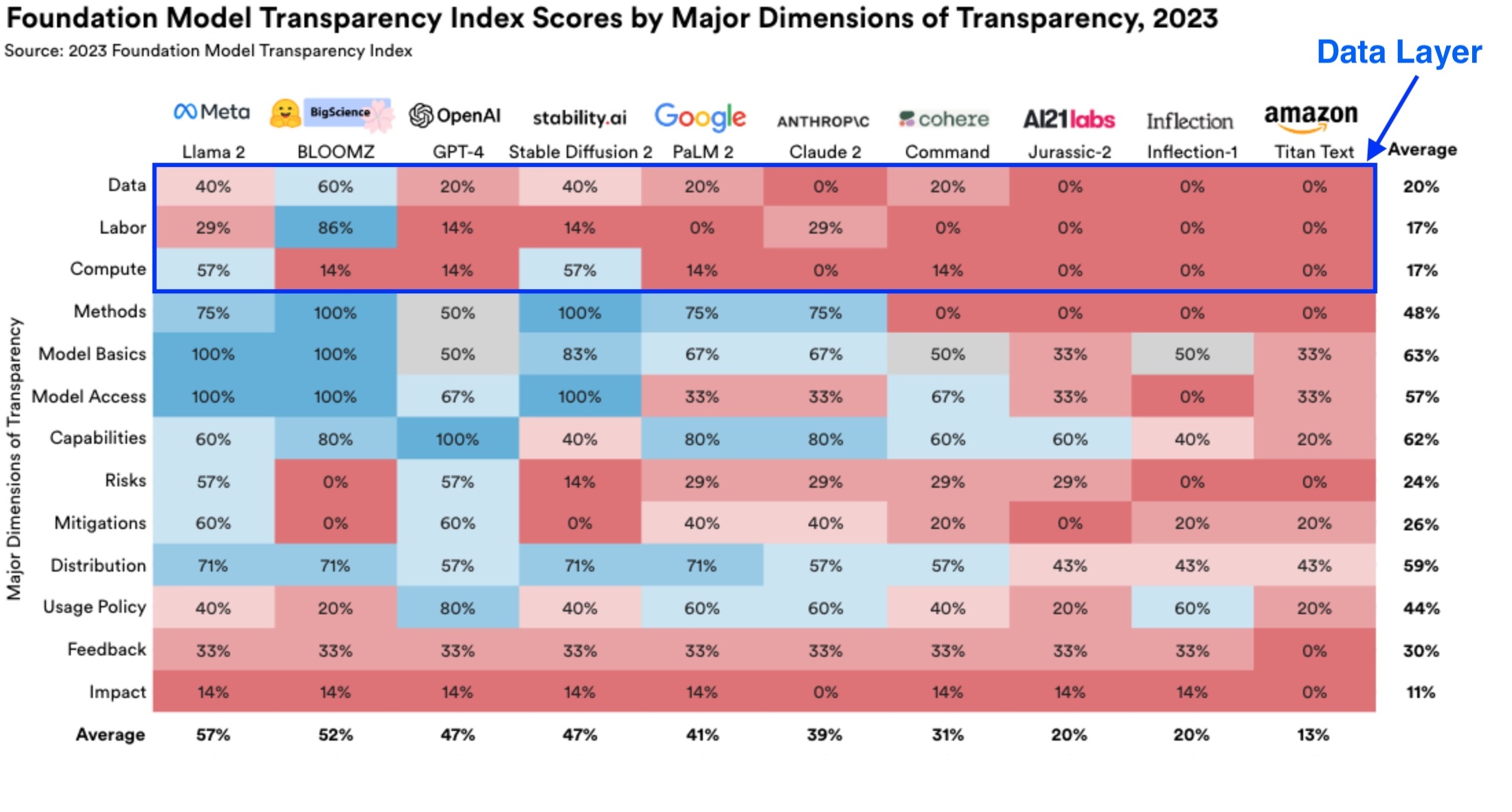}
    \caption{Figure from Bommanasi et al. (2023), in their Foundation Model Transparency Index paper comparing transparency across 10 key foundation models and 10 aspects of AI ecosystem transparency. Their ‘data layer’ includes data, labour and compute factors.
}
    \label{fig:enter-label}
\end{figure*}

Calls for improved transparency have emerged from various domains, and look increasingly likely to become an expectation - for example, in recent months, both the US \cite{the_white_house_fact_2024} and UK \cite{adams_algorithmic_2024} governments have rolled out requirements for public bodies to offer transparency into public sector AI use. While transparency can certainly not be considered a silver-bullet for addressing the ethical challenges associated with AI systems, it is a prerequisite for informed decision-making and other forms of intervention such as regulation and other forms of accountability \cite{nist_plan_2024}. 

\subsection{Where are we now?}

As discussed earlier in this section, a variety of best practices and standardised approaches for sharing transparency information have become increasingly well-established over the past several years. However, while recent studies have suggested that both data- and model- documentation practices are increasingly used, particularly among developer communities, evidence suggests that information supplied within documents is often limited and highly inconsistent \cite{liang_whats_2024, yang_navigating_2024}.

This is corroborated by recent research into transparency of the most popular AI datasets \cite{longpre_data_2023} and the Stanford research exploring AI lifecycle transparency within high profile foundation models (\cite{bommasani_foundation_2023-1}, see figure 1), which have demonstrated low availability of data transparency information in practice. Further, a Microsoft study demonstrated the limited data transparency information available to UX designers seeking to responsibly implement pre-trained models \cite{liao_designerly_2023}.

Still, there remains a need to consider how such findings generalise across a wider range of AI system types and applications, given the increasing deployment of AI products  across sectors and fields of practice. This is important, for example, for differentiating sectors, types of AI systems, and types of organisation in terms of data transparency information they provide, potential barriers for sharing this information, and particular targets for intervention. In responding to this need, this research considers AI data transparency in the context of systems generating public concern as a more diverse, priority sample of AI systems to examine. 

\section{Methodology}

\subsection{Research question}

In this research, we explore the question: what is the state of data transparency in AI systems that have generated public concern? 

\subsection{Data: AI incidents for identifying AI systems causing public concern}
The practice of recording AI incidents has emerged mainly with initiatives from civil society in recent years, primarily with the aim of helping to prevent recurrence of issues \cite{turri_why_2023}. ‘Incidents databases’ largely do not include transparency information about ‘upstream’ aspects of an AI system (how it was built and how that has led to the incident), but rather focus on impacts and ‘observable’ aspects of AI systems \cite{turri_why_2023}, and as such largely do not cover data transparency information. Still, the records of AI incidents can be used as a basis for exploration of how AI is used in the public domain, and the nature of concerns arising. 

\begin{figure*}[h!]
    \centering
    \includegraphics[scale=0.4]{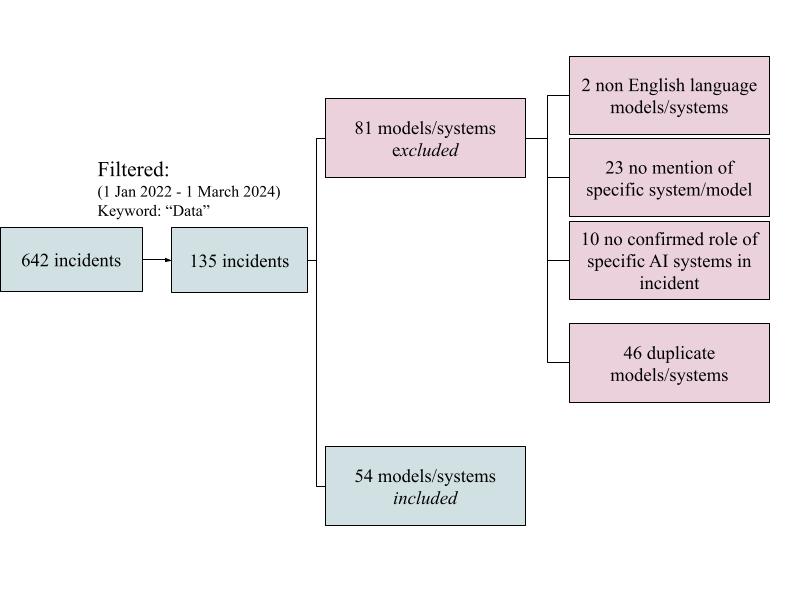}
    \caption{Overview of methodology}
    \label{fig:enter-label}
\end{figure*}

In this research we made use of the AI Incidents Database (AIID) released in 2020 by the Partnership on AI \cite{mcgregor_when_2020}. The database is developed on the basis of crowdsourcing incidents from the public which are then audited by administrators, who curate the incident reports based on information available in public news reports. Although there are multiple databases focused on publicly recording AI incidents, we selected the AIID due to easier interpretability of the sample and accessibility of the data. We discuss some limitations of this sample at the end of this paper.  

\subsection{Methods}
An overview of the methods used is available in Figure 2. 

\subsubsection{Filtering sample of AI systems}
This stage of the research was implemented by the first author of the paper. The AIID database contained 642 incidents as of early March 2024. We filtered these for recent incidents that took place between January 2022 and March 2024, to increase the likelihood of transparency information remaining available in findable formats. We used the AIID platform's search function to identify incident reports that mentioned ‘Data’ to limit our sample to those incidents that discuss any issues associated with data. This generated a list of 135 incidents for analysis.  

We used the AIID ‘incident report’ on the platform to identify the system and/or model at the centre of each AI incident. The exclusion criteria were all duplicate systems/models (46), non-English language systems (2), incidents where the issue was not clearly associated with the AI system (10), and any incidents that did not name a system/model/function of model such that it was difficult to include this information or a clear proxy (23). As such, we included 54 systems with a clear system name and function OR model name in our assessment.  

\subsubsection{Analysing AI data transparency}
In assessing ‘data transparency’, we used the list of indicators under the ‘Data’ section (one of ten transparency areas in their ‘ecosystem approach’ to AI transparency, see figure 1) in the Stanford Foundation Model Transparency Index (FMTI), developed from their review of scientific literature and public discourse on what AI transparency is needed for “public accountability” (p. 28, \cite{bommasani_foundation_2023-1}) among foundation models. This is defined as information needed to understand decision-making in building AI systems. Their research aimed to identify publicly available information about “data used to build the model” (in training, pre-testing, evaluation etc.). We have made use of their search protocol, openly documented online, \cite{bommasani_foundation_2023} to identify data transparency information, carefully documenting the search process and scoring each model by the data transparency indicator. However, this research faced resource constraints limiting certain aspects of the search protocol approach - for example, while the Stanford researchers validated their findings by contacting companies included in the sample, we used only publicly available documentation on the internet. In cases where the model is an ‘integration’ of other models (e.g., those built on ChatGPT such as in the case of FullPath’s GPT model) we specifically considered data used in this integration, not the original model. While relevant, we did not include ‘data labour’ and ‘compute’ indicators (see figure 1) in this research, also due to resource constraints. Future research on data transparency could explore such factors.

The indicators for ‘data transparency’ are: Data size; Data sources; Data creators; Data source selection; Data curation; Data augmentation; Harmful data filtration; Copyrighted data; Data licence; Personal information in data \cite{bommasani_fmti-indicators_2023}. 

\begin{figure*} [h!]
    \centering
    \includegraphics[scale =0.75]{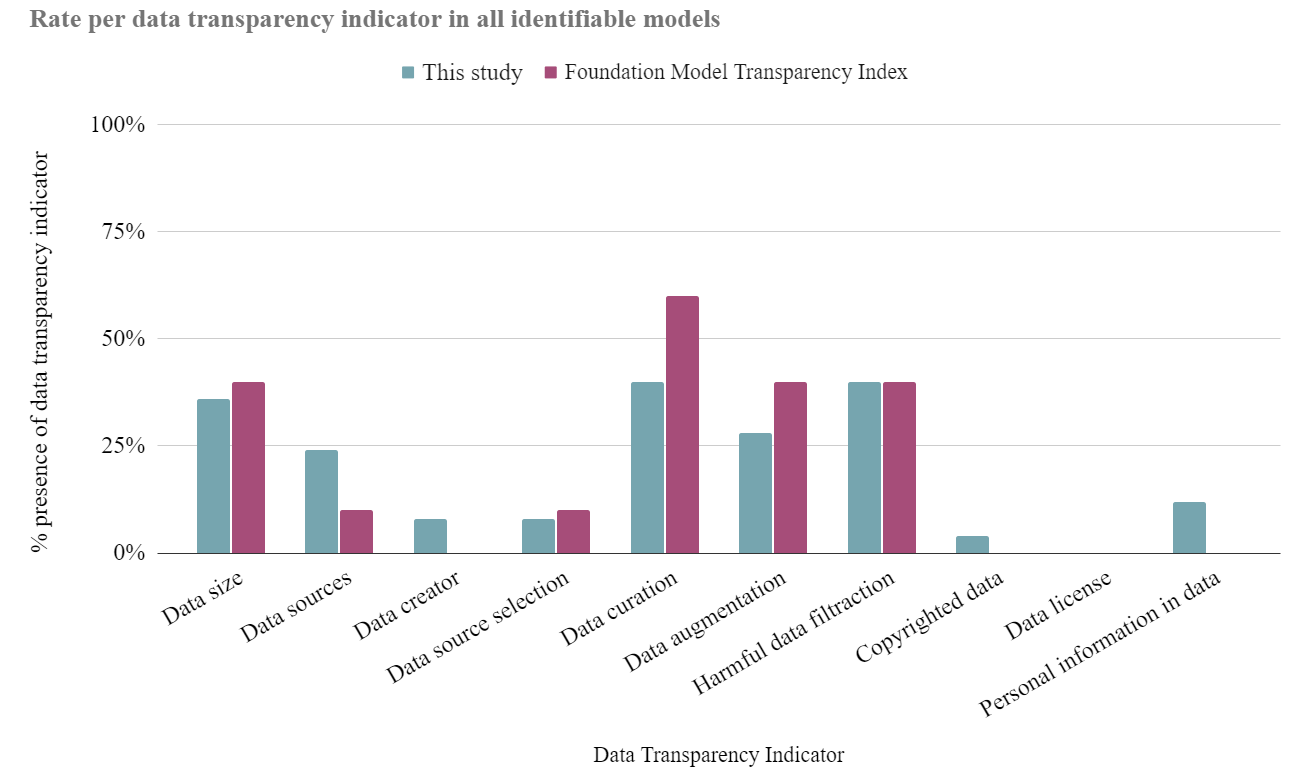}
    \caption{AI models scoring a point for each data indicator in this research (n=25) in comparison to the findings of the Foundation Model Transparency Index (n=10) (\cite{bommasani_foundation_2023}}
    \label{fig:enter-label}
\end{figure*}

\begin{figure*} [h!]
    \centering
    \includegraphics[scale=0.82]{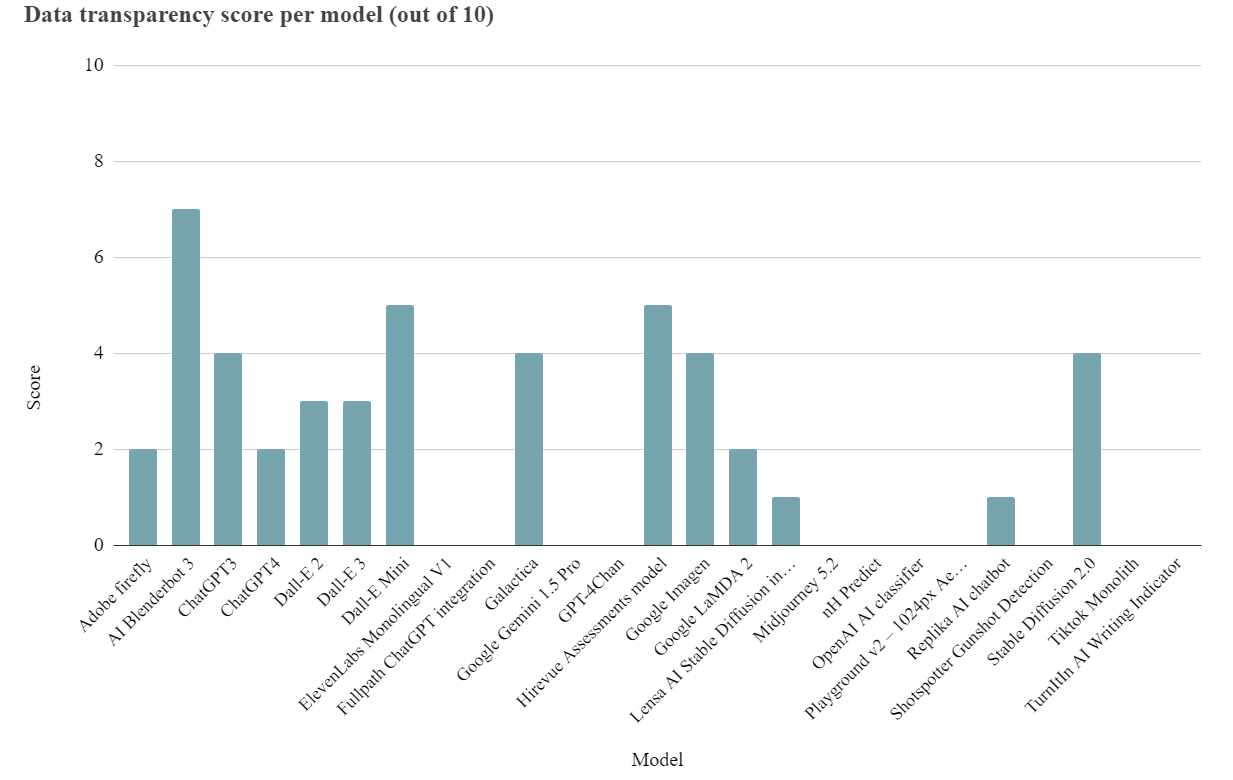}
    \caption{Comparing number of data transparency indicators across all AI models analysed}
    \label{fig:enter-label}
\end{figure*}

\begin{table*}[h!]
\begin{small}

\centering

\begin{tabular}{@{}lll@{}}
\toprule
\textbf{Function}   & \textbf{No.} & \textbf{System name(s)}                                         \\ \midrule
Chatbot &
  8 &
  \begin{tabular}[c]{@{}l@{}}AI Blenderbot 3 ; LaMDA 2 ; ChatGPT3 ; ChatGPT4 ; Fullpath ChatGPT integration ; Replika AI \\ chatbot ; GPT-4Chan ; Google Gemini 1.5 Pro\end{tabular} \\ \midrule
Image generation &
  7 &
  \begin{tabular}[c]{@{}l@{}}Dall-E 2 ; Dall-E 3 ; Dall-E Mini ; Google Imagen ; Lensa ; AI image generator ; \\ Midjourney ; Stable Diffusion 2.0\end{tabular} \\ \midrule
Image editing             & 2            & Adobe Firefly Image 2 ; Playground v2 – 1024px Aesthetic Model \\ \midrule
Generative AI identification      & 2            & OpenAI AI classifier ; TurnItIn AI Writing Detection                           \\ \midrule
Insurance decisions       & 1            & \multicolumn{1}{l}{nH Predict}                               \\ \midrule
Interview assessment & 1            & Hirevue Assessments model                         \\ \midrule
Recommendations  & 1            & Tiktok Monolith                                \\ \midrule
Sound identification         & 1            & Shotspotter                    \\ \midrule
Voice generation    & 1            & ElevenLabs Monolingual V1                                                       \\ \midrule
Writing papers   & 1            & Meta Galactica                             \\ \bottomrule
\end{tabular}
\caption{Systems and functions (model information identified) (n=25)}\label{tab:first}
\bigskip

\begin{tabular}{@{}lll@{}}
\toprule
\textbf{Function}   & \textbf{No.} & \textbf{System name(s)}                                         \\ \midrule
Autonomous driving &
  7 &
  \begin{tabular}[c]{@{}l@{}}XPeng P7 ; Cruise autonomous vehicle technology ; Serve Robotics ; Tesla 'full self driving' \\ software ; Tesla Model 3 Sedan Autopilot ; Tesla Model 9 Autopilot mode ; TuSimple \\ autonomous vehicle technology\end{tabular} \\ \midrule
Facial recognition technology &
  5 &
  Amazon Rekognition ; CBP One ; Clearview AI facial recognition software ; Hikvision ; Pimeyes \\ \midrule
Recommendations &
  5 &
  \begin{tabular}[c]{@{}l@{}}Amazon recommended purchases ; Facebook job ads algorithm ; \\ Google docs 'inclusive language' function ; Google search ; Instagram recommender algorithms\end{tabular} \\ \midrule
Chatbot             & 2            & Microsoft Co-Pilot Chatbot ; Tessa chatbot ; Chai Eliza chatbot \\ \midrule
CSAM detection      & 1            & Google Photos image analysis software                           \\ \midrule
Image editing       & 1            & \multicolumn{1}{l}{ClothOff app}                               \\ \midrule
Pollution detection & 1            & Toronto AIPM system - Virtual beach 3.0                         \\ \midrule
Purchase detection  & 1            & Amazon Fresh detection algorithm                                \\ \midrule
Translation         & 1            & Google Lens camera-based translation feature                    \\ \midrule
Video generation    & 1            & Synthesia                                                       \\ \midrule
Voice recognition   & 1            & Centrelink voice recognition system                             \\ \bottomrule
\end{tabular}
\caption{Systems and functions (no model information identified) (n=29)}\label{tab:second}
\end{small}

\end{table*}

\subsubsection{Applying the search protocol}
To apply the search protocol required identification of the model underlying the AI system’s function. As discussed in section 2.3, an AI system is made up of the inputs (data) and the mechanisms (‘models’) that generate the output - and an AI system’s outputs can be generated from multiple models, or just one \cite{meta_system_2022}. To understand the data used to build a system, it was important to identify the particular model(s) underlying a given function. Where the specific model was not named in the incident reports, for each system we searched “COMPANY NAME” +”SYSTEM NAME/FUNCTION” + “model transparency”/”technical paper”/”model” and explored the resulting findings to select a model that was released within the relevant time period for the incident. For those where we could not identify a clear input-output mechanism/model underlying the system/function (we included those where this was not explicitly called a 'model'), we considered that we were not able to apply the search protocol outlined. We discuss the challenges in identifying model transparency information in sections 4.1 and 5.1.

To build on the research protocol and identify any remaining information, we search for key transparency information using further searches for “MODEL NAME” + “transparency”/“data transparency”/”model transparency”/”technical paper” and recorded any key transparency sources, ensuring we searched in these for data transparency information. 

\section{Findings} 

\subsection{Identifying models and model transparency information} 
For just over half of the 54 systems, we were not able to identify information about the model architecture/specific models or algorithms associated with the concern, such that we could apply the data transparency search protocol. 29 systems/products did not offer identifiable model information, while 25 did and were analysed further for data transparency indicators. 

The functions of those systems with identifiable model information (see table 1) were mostly associated with chatbots (n=8) or image generation (n=7), while the functions of those systems without identifiable model information (see table 2) were primarily associated with autonomous driving (n=7), facial recognition (n=5) and recommendation models (n=5).

For example, it is difficult to unpack how notoriously opaque \cite{stray_show_2021} recommender systems by big tech organisations work, likely a consequence of both the involvement of many different algorithms/models playing a role within the system/function at hand \cite{lada_how_2021}, and a desire by companies to keep the workings of their systems private to protect company interests \cite{luria_this_2023}. Even where Meta, for example, has shared ‘system cards’ covering a large portion of their AI-driven recommender systems \cite{meta_our_2023}, these tend to highlight that ‘multiple models’ are used to generate a particular function, and they do not name or offer more in depth information about these and the data used to develop them. As such, while it is often clear the general types of data these systems rely upon (customer purchasing history, browsing history etc.), it is difficult to identify a particular ‘input-output’ model or algorithm that the incident was associated with, in order to carry out the data transparency search protocol in this research. We identified similar challenges for identifying the ‘models’ at stake in facial recognition and autonomous driving systems among others (see Table 2) without specialist knowledge of the systems.

\subsection{Assessing data transparency within identifiable models}

For those systems where we were able to apply the search protocol (i.e., those with identifiable model architectures/algorithms), we found that systems scored on average 2 of the 10 data transparency indicators, with 9 (40 percent) of the models scoring zero points. Of these, particular data indicators were more likely to be included (figure 3). Looking in particular at the nature of data transparency information available, the most commonly available information was about data curation and harmful data filtration (40 percent prevalence), data size (36 percent) and data sources (24 percent). Points were awarded in cases where some significant information was available, even if this was not a comprehensive description of the indicator, per the FMTI protocol. None of the models examined gave information about data licensing and very few gave information about personal information in data (12 percent), data creators (8 percent), and data source selection (8 percent) and copyrighted data (4 percent). Similarly, the Stanford FMTI researchers found that none of the 10 foundation models they assessed gave transparency information about data creators, data licensing, copyrighted data, nor personal information within the data used  \cite{bommasani_foundation_2023}). 

\section{Discussion}

\subsection{Diverse challenges and ‘path-dependency’ for data transparency across AI systems}

AI exists in society in many different forms. The findings of this research have demonstrated that the AI systems generating public concern are far from homogenous\footnote{indeed they will be, as discussed in section 6, far more diverse than those represented in this sample.}, and with this come diverse challenges in understanding how data has been used to build these AI systems.

Our findings demonstrate a ‘hierarchy of transparency’ or ‘path-dependency’, where it is difficult to investigate upstream elements of a system without understanding the basic architectures of a system or product. In this research, the vast majority of cases where AI models were identifiable and documented were generative AI-based systems (see Table 1). With the high profile technologies and models behind products for functions such as text and image generation, model names appear to be commonly reported in news, blogs and other materials online. The protocol used in this research to identify AI data transparency was taken from an existing method designed for generative AI/foundation models. To better understand AI data transparency for public accountability, future research should attempt to account for the diverse challenges for understanding how data has been used to build different AI systems, and the different ‘levels’ at which transparency is prevented.

Further, we agree with the various studies concluding the importance of ‘holistic’ \cite{heger_understanding_2022, micheli_landscape_2023, brereton_role_2023} approaches to AI transparency that help to understand the AI lifecycle - and emphasise the importance of embedding data transparency effectively in such approaches. A commonly touted solution is to embed data cards within model cards, and likewise model-level documentation can be embedded within system-level documentation (which take a more holistic view of the product and how it works). 

\subsection{Lack of data transparency of AI systems and models behind AI incidents}

Largely, our findings suggest that the public availability of data transparency information is very limited across the systems identified as associated with AI incidents. Where it has been possible to identify particular models underlying AI incidents, the findings strikingly corroborate those of the Stanford Foundation Model Transparency Index, supporting their evidence generated across a sample of 10 foundation models with a wider range of 25 AI models.

It is interesting to note that the majority of those models scoring more than zero points were developed by ‘big tech’ companies (see figure 4). While the evidence in this research is insufficient to substantiate such a conclusion, it may be interesting to explore whether smaller companies (including those stewarding generative AI models) are even less likely to offer data transparency information. The vast majority of incidents within the AI Incidents Database are indeed associated with big tech providers, highlighted as a limitation of the sample used in this report, and we suggest that future research could consider the disparities in data transparency across different types of organisations/providers. 

The data transparency indicators designed by the Stanford FMTI and used in this research were designed based on evidence of public discourse and existing scientific literature on the need for transparency about foundation models. Particularly where there are very few or no models scoring on an indicator, it is important to reconsider how realistic and important these transparency elements are for developers to provide, across different system types. For example, we only identified  transparency information addressing whether copyrighted data had been included in the model’s development for one of the models included in this sample (Adobe’s Firefly) - and none of those assessed in the FMTI scored this point. This should be interpreted in the context of major concerns and legal challenges \cite{vincent_ai_2023} raised over the use of copyrighted data within AI systems, and particularly generative AI systems, leading to lobbying for legal requirements on companies to disclose the copyright status of their data. While almost none of those systems identified in this research respond to such demand, there is clear appetite among a diversity of stakeholders for copyright information to emerge \cite{david_news_2023, david_ai_2023}. Similarly, the range of AI incidents focused on concerns surrounding harmful data inclusion, data biases, and the disclosure of personal data, for example, suggest that such indicators are applicable across a wide range of AI systems. The absence of this information raises questions about the reason such information is not shared - barriers to sharing information about the data in AI systems commonly highlighted \cite{felzmann_towards_2020} include proprietary reasons, cost, and impracticality, although there appears to be a lack of consensus regarding the extent of such barriers and their solutions.

Reconsideration of the applicability of the data transparency indicators used in this research across different types of stakeholders’ needs, and for understanding different types of systems (and indeed, the types of AI incidents they are commonly associated with) may help to develop further approaches for monitoring the status of AI data transparency. As is highlighted as an objective by the FMTI researchers \cite{bommasani_foundation_2023-1}, this may help, in turn, to communicate ‘best practice’, clear expectations and highlight gaps effectively to developers. 

\subsection{AI documentation practices and addressing stakeholder needs}

As discussed in sections 1 and 2, AI documentation is the topic of significant attention, responding to the demand for transparency across many different stakeholder types. Much of the transparency information identified in this research was found in model cards and technical papers, suggesting the importance of the increasing expectation to supply these transparency documents for data transparency. Still, as demonstrated by the generally low scores among AI models assessed, the presence of model or system transparency documentation does not necessarily lead to presence of all desired transparency information. 

Our findings have supported numerous other studies \cite{liang_advances_2022, yang_navigating_2024} demonstrating a need to increase the consistency of use of AI transparency documentation. It is important to consider how to ensure that data- and model- level transparency approaches lead to information being available to both technical and non-technical audiences seeking to understand the role of data within the systems and their downstream impacts, and therefore how these are incorporated in system-level transparency approaches, and also whether the information supplied is appropriate/sufficient. Still, we also highlight the importance of a continued research agenda to understand the efforts and barriers for producing this documentation, and how this can be best supported and encouraged. 

\section{Limitations of approach}
This sampling and coding in this research was undertaken by a single researcher. As such, the findings are indicative of the availability of information based on the FMTI search protocol, but may not reflect all available information. Additionally, due to resource constraints we were not able to contact AI system providers to offer an opportunity to dispute our findings, a step included in the FMTI research approach. Contacting the system providers may have led to further interesting insights in addition to validation of our findings.

Further, the AI incidents reported in news media platforms and recorded to AI incidents databases will be a limited representation of those AI systems causing public concerns/real world incidents, with biases towards ‘big tech’ concerns, those affecting Global North countries, and less marginalised groups who may be more likely to have issues affecting them gain global attention. Indeed, in this research we were limited to analysing the English language examples. As such, expansion of this research approach to identify data transparency among a broader range of incidents would be important to validate and build on the findings. 

Finally, the data transparency indicators used in this research protocol were designed specifically for foundation model transparency. While these indicators still appear broadly relevant across all types of AI systems, in section 5 we discuss the need to develop a protocol for assessing data transparency within a wider range of AI systems, including consideration of which indicators are most appropriate across diverse systems.

\section{Recommendations}

\subsubsection{Recommendation 1: Investigate further how data transparency empower users of transparency information}

In this research we have identified challenges in understanding the systems, models and datasets at stake in AI systems, and the importance of carefully considering how data transparency is embedded within AI system/ecosystem level transparency approaches. Particularly given the perceptions and presentation of data as ‘technical’ despite its clear relevance for understanding the impacts on individuals and technology users - and therefore many along its ‘supply chain’ for whom impacted individuals are their customers, clients etc. - it is important to consider the AI data transparency ‘information needs’ of different AI stakeholders more closely and how this varies across different contexts. In particular, given the most detailed data documentation approaches (dataset-level documentation) are aimed primarily at AI developers, we suggest the importance of considering how data transparency information for understanding AI systems and AI incidents can empower non-specialists and communities. Still, as discussed in recommendation 2, there remains a need to ensure simplicity and feasibility of transparency approaches, and to consideration of how to balance the needs of diverse stakeholders. 

\subsubsection{Recommendation 2: Better investigate the barriers and opportunities for effective and efficient sharing of data transparency information}

Common to discussions of data transparency are debates over the complexities of sharing data transparency information. Developers highlight concerns over privacy, security and intellectual property - and there are certainly questions in terms of explainability (are people able to understand the information they need), cost of generating transparency information, and incentives. 

By generating an improved understanding of the different challenges for organisations developing AI systems, it may be possible to identify approaches that navigate significant concerns, while preventing calls for transparency being undermined by more vague and high-level counter-arguments such as ‘privacy concerns’ in spite of real demand for information and where issues could be mitigated.

For example, exploration of the potential challenges and opportunities for automated documentation of data practices across different system types may support more feasible and standardised approaches \cite{thirumuruganathan_automated_2021}.

\subsubsection{Recommendation 3: Develop an AI data transparency index addressing a wide range of AI systems}

We view the ecosystem-level view of the Foundation Model Transparency Index as key to understanding the needs and gaps in transparency approaches for these particular models across all aspects of the AI ecosystem. However, given the particularly limited availability of transparency information on AI data, and the diverse challenges raised in this report for understanding the data used in AI systems, we suggest that it is important to build on the findings of the FMTI and this exploratory report to develop a systematic means of assessing AI data transparency across more diverse systems. A data transparency index may consider:
\begin{itemize}
  \item Both ‘big tech’ and smaller providers’ data transparency
  \item How data transparency varies across diverse applications of AI (in healthcare, environmental contexts, public sector etc.) - this may help to prompt an understanding of areas of best practice and areas of significant need and policy attention
  \item Data transparency indicators that empower different communities who represent societal interest in the implementation of AI systems (see recommendation 2)
  \item Transparency assessment of key datasets
\end{itemize}
We intend this research to offer a starting point for such efforts.

\section{Conclusion}

In all, the rapidly evolving landscape of AI data transparency has some way to go before the principle of transparency can be seen to be met. Our findings agree with the broad consensus that data documentation within AI systems remains largely ad hoc and mostly opaque to non-specialists and those outside of the companies creating the systems. We do, however, recognise the progress that has been made in recent years with increasing expectations for use of AI documentation practices such as data and model cards, and publishing ‘technical papers’, FAQs, blogs and more to improve public accountability and trust, and suggest that these have met particular success with improving the public accountability of foundation models (perhaps in response to the high level of concern about these systems). 

We further agree that the approach by the Stanford team developing the Foundation Model Transparency Index for generating an understanding of the AI transparency landscape is important for raising awareness of needs and gaps in transparency, and propose that deeper consideration of the AI \textit{data} transparency is needed to address the particular limitations for public accountability of AI systems.

\section{Acknowledgments}
This exploratory research was undertaken as part of the Open Data Institute's Data-Centric AI programme, for which we are grateful for funding by the Patrick J. McGovern Foundation. 

\bibliography{MAIN_TEXT_FILE}

\end{document}